\documentclass[
aps,
prb,
reprint,
superscriptaddress,
longbibliography
]{revtex4-2}
\usepackage{amsfonts}
\usepackage{mathrsfs}
\usepackage{amsmath}
\usepackage{color}
\usepackage{natbib}
\usepackage{graphicx}
\usepackage{bm}
\usepackage{amssymb}
\usepackage{xspace}
\usepackage{epstopdf}
\usepackage{dcolumn}
\usepackage{multirow}
\usepackage[colorlinks=true, letterpaper=true, pdfstartview=FitV, linkcolor=blue, citecolor=blue, urlcolor=blue]{hyperref}
\usepackage{wrapfig}

\makeatletter

\newcommand{\Rmnum}[1]{\expandafter\@slowromancap\romannumeral #1@}
\makeatother

\begin{document}

\title{Pure Spin Photocurrent in Altermagnetic Photovoltaic Battery}

\author{Qiang Li}
\affiliation{Department of Physics, Hubei Minzu University, Enshi 445000, P. R. China}
\affiliation{Science, Mathematics and Technology (SMT) Cluster, Singapore University of Technology and Design, Singapore 487372, Singapore}

\author{Shibo Fang}
\email{shibo\_fang@sutd.edu.sg}
\affiliation{Science, Mathematics and Technology (SMT) Cluster, Singapore University of Technology and Design, Singapore 487372, Singapore}

\author{Zongmeng Yang}
\affiliation{Science, Mathematics and Technology (SMT) Cluster, Singapore University of Technology and Design, Singapore 487372, Singapore}
\affiliation{State Key Laboratory for Mesoscopic Physics and School of Physics, Peking University, Beijing 100871, P. R. China}

\author{Xingyue Yang}
\affiliation{Science, Mathematics and Technology (SMT) Cluster, Singapore University of Technology and Design, Singapore 487372, Singapore}
\affiliation{State Key Laboratory for Mesoscopic Physics and School of Physics, Peking University, Beijing 100871, P. R. China}

\author{Jianhua Wang}
\affiliation{Science, Mathematics and Technology (SMT) Cluster, Singapore University of Technology and Design, Singapore 487372, Singapore}
\affiliation{School of Materials Science and Engineering, Tiangong University, Tianjin 300387, China}

\author{Rui Peng}
\affiliation{School of Physics and Optical Engineering, Zhejiang University of Technology, Hangzhou 310023, China}

\author{Lin Zhu}
\affiliation{Science, Mathematics and Technology (SMT) Cluster, Singapore University of Technology and Design, Singapore 487372, Singapore}
\affiliation{College of Chemistry and Pharmacy, Northwest A\&F University, Yangling, Shaanxi 712100, P. R. China}

\author{Shuhua Wang}
\affiliation{Energy Research Institute @ Nanyang Technological University, ERI@N, Interdisciplinary Graduate School, Nanyang Technological University, 50 Nanyang Avenue, Singapore 639798, Singapore}

\author{Dexing Liu}
\affiliation{School of Science and Engineering, The Chinese University of Hong Kong, Shenzhen, Shenzhen 518172, China}

\author{Min Zhang}
\affiliation{School of Science and Engineering, The Chinese University of Hong Kong, Shenzhen, Shenzhen 518172, China}

\author{Dahua Ren}
\affiliation{Department of Physics, Hubei Minzu University, Enshi 445000, P. R. China}

\author{Mai Zhang}
\affiliation{The School of Electronic Information, Northwest University, Xi'an 710127, People’s Republic of China}

\author{Han Zhang}
\affiliation{The School of Electronic Information, Northwest University, Xi'an 710127, People’s Republic of China}

\author{Yee Sin Ang}
\email{yeesin\_ang@sutd.edu.sg}
\affiliation{Science, Mathematics and Technology (SMT) Cluster, Singapore University of Technology and Design, Singapore 487372, Singapore}

\begin{abstract}
Altermagnets, featuring momentum-dependent spin splitting without net magnetization, provide a promising platform for spintronic functionalities beyond conventional ferromagnets and antiferromagnets. Here, we propose an altermagnetic spin photovoltaic battery consisting of a nonmagnetic semiconducting layer sandwiched between two altermagnetic electrodes. Using first-principles quantum-transport simulations, we show that a V$_2$Te$_2$O/ZnSe/V$_2$Te$_2$O junction supports a pure spin photocurrent for opposite N\'eel vectors in the two altermagnetic electrodes, with spin-up and spin-down photocurrents equal in magnitude and opposite in sign. The effect persists under both linearly and circularly polarized light and remains tunable with photon energy and polarization angle. Our results establish a realistic route toward light-driven pure spin-current generation in altermagnetic junctions.
\end{abstract}

\maketitle

\section{Introduction}

Spin-polarized charge currents and pure spin currents are two foundational pillars of spintronics \cite{Zutic2004Spintronics,altz2018Antiferromagnetic}. Spin-polarized currents form the basis of conventional spintronic architectures, such as those built from magnetic materials and magnetic tunnel junctions (MTJs) \cite{Wu2023FexGeX2MTJ,Wu2023MnSe2MTJ,guo2026sliding,bai2026ferroelectrics}, while pure spin currents offer a route toward dissipationless information transfer without net charge motion \cite{Fang2024LightAssistedNeel,Fei2021SpinCircularPhotogalvanic}. In MTJs, for example, spin-polarized currents exert spin-transfer torques, whereas pure spin currents generated by spin-orbit coupling give rise to spin-orbit torques \cite{Shao2023NeelSpinCurrents,Shao2024AFTunnelJunctions,Song2021SOTReview}. However, in conventional spintronic platforms, pure spin currents are often generated through spin-orbit conversion in heavy metals with strong spin-orbit coupling, which restricts material choices and can introduce additional Joule dissipation as well as interfacial losses \cite{nguyen2024recent,han2021spin}.

Beyond electrical manipulation, light-matter interactions offer an alternative way to generate and manipulate spin-polarized and pure spin currents \cite{Dai2023BulkPV,Fei2021SwitchableSpinPhotocurrent,Fang2024SpinPhotogalvanic,Zhu2025MagneticGeometry,yang2025nonlinear,g85x-rgxm}. Their large excitation energies, sub-picosecond temporal resolution, and non-contact nature make them attractive for integration with ultrafast spintronic devices \cite{Fei2020PhotogalvanicAxion,Li2024PhotodetectorBPBi,Tengdin2020LightInducedSpinTransfer,Ang2010Nonlinear}. Importantly, the characteristic timescales of optical excitations naturally match the operational bandwidth of antiferromagnetic (AFM) spintronics, where N\'eel-vector switching frequencies can reach the terahertz regime \cite{Yang2024MultistateMTJ,Fu2025FloquetTriplet,Kotegawa2024AnomalousHallfElectron,Xu2021BulkSpinPV,zhang2026deterministic,wang2026electric}. Femtosecond pulses can generate ballistic and shift photocurrents along specific crystallographic directions, and the resulting photocurrents can exert sufficiently strong torques to reorient the N\'eel order \cite{Zhou2025SpinTorqueContrast,Fang2024QuantumGeometryAltermagnets}. These features highlight the strong potential of light as an efficient and versatile route for generating and controlling spin currents.

Altermagnets (AMs) have recently emerged as an unconventional class of antiferromagnets hosting symmetry-protected, nonrelativistic spin splitting \cite{jungwirth2026symmetry,guo2025spin,duan2025antiferroelectric,kqy8-myz1,vsmejkal2022emerging}. Their sizable momentum-dependent spin splitting, which can arise even without spin-orbit coupling, endows them with ultrafast and stray-field-free spin dynamics while enabling efficient spin selectivity \cite{Jiang2025MetallicdWaveAltermagnet,Song2025AltermagnetsReview,Ma2021PiezomagnetismAntiferromagnet,Hu2025CPairedSpinMomentum,fang2026alterelectricityelectricalanaloguealtermagnetism,peng2025ferroelastic}. Recent studies have begun to explore the optoelectronic properties of AMs, revealing a range of light-driven spin phenomena, including nonlinear spin photovoltaic effects, symmetry-dependent magnetization, and optical responses distinct from those of conventional antiferromagnets \cite{g85x-rgxm,dong2025crystal,Zhou2025SpinTorqueContrast}. At the same time, AM-based device concepts have been developed predominantly in the context of electronic transport, such as tunnel junctions and Hall-bar geometries \cite{Fang2025Edgetronics,Wang2025PentagonalAltermagnets,Ezawa2025BPVAltermagnets,Choi2024MetaOptics,Han2024NeelVectorSwitching,Luo2025TMRNoncollinearAFM,Yang2025TMRAltermagnets,yang2026altermagnetic,Yang2026AMTJ}. While optoelectronic implementations have also started to emerge, their development remains at an early stage, and further exploration of AM-based photovoltaic device architectures, particularly those analogous to conventional heterostructure photodetectors, is of considerable interest \cite{Xie20182DSpinBattery,Fang2023BidirectionalPhotoresponse,liu2026spin}.

In this work, we propose an altermagnetic spin photovoltaic battery based on a metal-semiconductor-metal junction, in which a photoactive nonmagnetic semiconductor is sandwiched between two altermagnetic electrodes. The operating principle relies on spin-selective photocurrent extraction imposed by the altermagnetic electrodes. For opposite N\'eel vectors in the two electrodes, the device generates a pure spin photocurrent, with spin-up and spin-down photocurrents equal in magnitude and opposite in sign. Using first-principles quantum-transport simulations, we demonstrate this mechanism in a V$_2$Te$_2$O/ZnSe/V$_2$Te$_2$O device and characterize its dependence on photon energy and light polarization angle. These results highlight the potential of altermagnetic junctions for spin-optoelectronic applications.

\begin{figure}[t]
  \centering
  \includegraphics[width=\columnwidth]{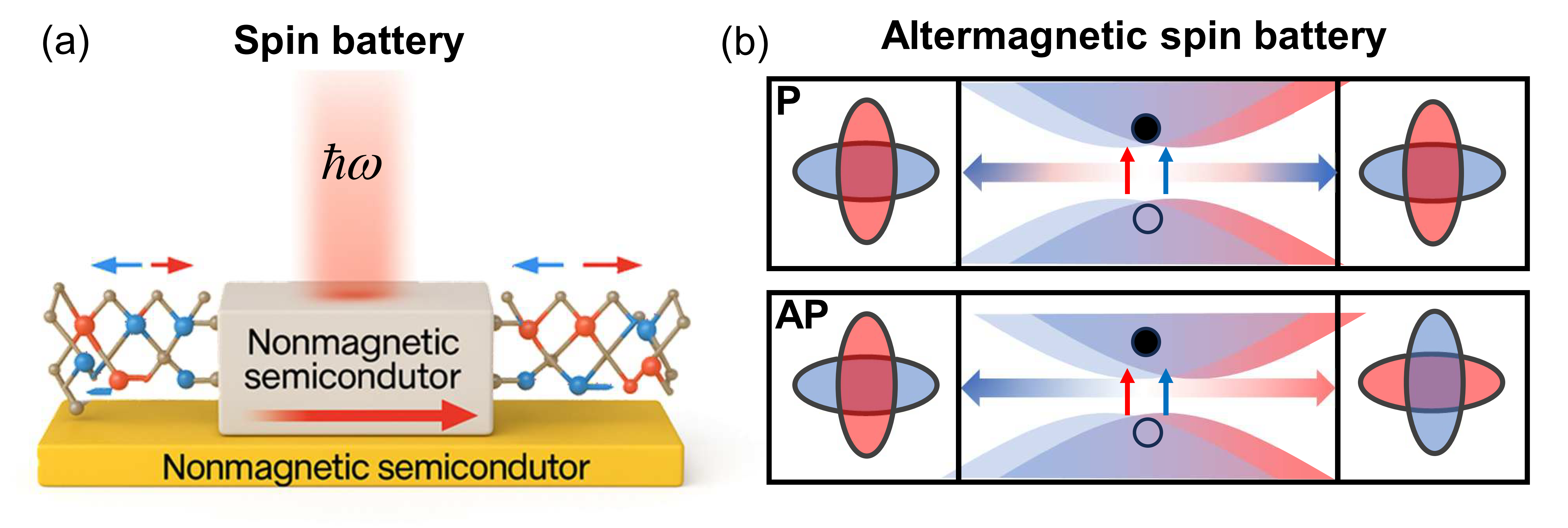}
  \caption{Schematic illustration of the altermagnetic spin photovoltaic battery. (a) Device configuration, consisting of a nonmagnetic semiconducting layer sandwiched between two altermagnetic electrodes under illumination. (b) Operating mechanism for parallel (P) and antiparallel (AP) N\'eel configurations of the electrodes. Red and blue arrows denote spin-up- and spin-down-polarized currents, respectively.}
  \label{fig:fig1}
\end{figure}

\section{Method}
First-principles calculations are performed within the framework of density-functional theory (DFT). The exchange-correlation functional is described using the generalized gradient approximation (GGA) in the Perdew-Burke-Ernzerhof (PBE) form. Structural relaxations of the monolayer two-dimensional materials are carried out using projector-augmented-wave (PAW) pseudopotentials and a plane-wave basis, as implemented in the Vienna \textit{ab initio} simulation package (VASP)~\cite{Kresse1999PAW}. A plane-wave cutoff energy of 500 eV is adopted, and the Brillouin zone is sampled using a $12 \times 12 \times 1$ Monkhorst-Pack \textit{k}-point mesh. The atomic structures are relaxed until the residual force on each atom is smaller than $10^{-2}$ eV/\AA, and the total-energy convergence criterion is set to $10^{-5}$ eV.
Based on the optimized structures, the electronic band structures and quantum-transport calculations are further carried out using the linear combination of atomic orbitals (LCAO) basis as implemented in QuantumATK \cite{Smidstrup2020QuantumATK,Smidstrup2017GFSurface}. Electron correlation on the V $3d$ orbitals is treated within the GGA+$U$ scheme, where an effective Hubbard parameter of $U_{\mathrm{eff}} = 3.1$ eV is adopted.

The spin photocurrent is calculated using density-functional theory combined with the nonequilibrium Green's function (DFT+NEGF) method. An electronic temperature of 300 K and a density mesh cutoff of 105 Hartree are adopted in the photocurrent calculations. For the self-consistent calculations, a $50 \times 1 \times 235$ \textit{k}-point mesh is used for the source lead, while a $50 \times 1 \times 1$ mesh is used for the transport direction. Here, the $x$, $y$, and $z$ directions correspond to the periodic, vacuum, and transport directions, respectively. Periodic boundary conditions are applied along the $x$ direction, whereas Neumann and Dirichlet boundary conditions are imposed along the $y$ and $z$ directions, respectively.

The devices are simulated in a two-probe configuration within DFT+NEGF, as depicted in Fig.~\ref{fig:fig1}(a) \cite{Smidstrup2020QuantumATK,Stradi2016MetalSemiconductorInterface,vanSetten2018PseudoDojo}. The illuminated region of the device is exposed to incident light, and the photocurrent flowing into the left electrode is computed, as illustrated in Fig.~\ref{fig:fig1}(b). In the two-probe structure, the photocurrent is written as \cite{Chen2012Photocurrent,Henrickson2002PhotocurrentModeling,Zhang2014ValleyPolarized,liu2025shot}
\begin{equation}
I_L^{\mathrm{ph}}
=
\frac{e}{\hbar}
\int
\frac{1}{2\pi}
\sum_k
\mathrm{Tr}
\left\{
i\Gamma_L(E,k)\left[1-f_L\right]G^{<}
+
f_L G^{>}
\right\}
\, dE .
\end{equation}
Here, $L$, $e$, and $\hbar$ denote the left lead, electron charge, and reduced Planck constant, respectively. $\Gamma_L=i(\Sigma_L^r-\Sigma_L^a)$ is the linewidth function denoting the coupling between the central scattering region and the left lead, and $\Sigma_L^r=[\Sigma_L^a]^{\dagger}$ is the retarded self-energy due to the left lead. $f_L(E)$ is the Fermi-Dirac distribution function of the left lead. $G^{</>}=G_0^r(\Sigma_L^{</>}+\Sigma_R^{</>}+\Sigma_{\mathrm{ph}}^{</>})G_0^a$ is the lesser/greater Green's function including electron-photon interaction, where $G_0^{r/a}$ is the retarded/advanced noninteracting Green's function and $\Sigma_{L,R,\mathrm{ph}}^{</>}$ are the corresponding self-energies. A vacuum gap of approximately 30 \AA\ is employed perpendicular to the transport direction.

\section{Results}

\subsection{Altermagnetic spin photovoltaic battery}
The proposed AM spin photovoltaic battery consists of two altermagnetic metallic electrodes and a nonmagnetic, inversion-symmetric semiconductor as the central region, as shown in Fig.~\ref{fig:fig1}(a). When the two electrodes have the same N\'eel order (parallel, P state), the device preserves spatial inversion symmetry and no photocurrent is generated under illumination. By contrast, when the two electrodes have opposite N\'eel orders (antiparallel, AP state), the inversion symmetry is broken and the device supports a pure spin photocurrent with $I_{\uparrow}=-I_{\downarrow}$. As a result, the charge photocurrent vanishes, whereas a finite spin current is generated. In this mechanism, photoexcited carriers in the semiconductor are extracted through spin-selective transport channels imposed by the two altermagnetic electrodes, giving rise to two oppositely polarized photocurrent components of equal magnitude, as illustrated in Fig.~\ref{fig:fig1}(b).

Here, we adopt a conceptually simple spin-photovoltaic architecture based on altermagnetic electrodes, which is fundamentally different from several existing routes to pure spin-current generation. Compared with nonlinear quantum-transport proposals based on intrinsic or engineered $\mathcal{PT}$-symmetric antiferromagnets, the present design does not rely on highly artificial structural antisymmetry, such as perfectly antisymmetric perforations or adsorption patterns, and is therefore physically more transparent and potentially more robust experimentally \cite{Xie20182DSpinBattery}. It also differs from conventional ferromagnetic spin photovoltaic batteries, which typically generate spin-polarized charge currents with a finite net charge component \cite{Xie20182DSpinBattery}. In contrast, the altermagnetic device directly produces a pure spin photocurrent without an accompanying charge current. Relative to conventional antiferromagnetic platforms, altermagnetic electrodes further combine zero net magnetization with intrinsic momentum-dependent spin splitting, thereby providing a more direct route toward spin-selective photocarrier extraction. These features make the AM spin photovoltaic battery a simple and promising platform for light-driven pure spin-current generation.

\subsection{Material realization}

\begin{figure}[t]
  \centering
  \includegraphics[width=\columnwidth]{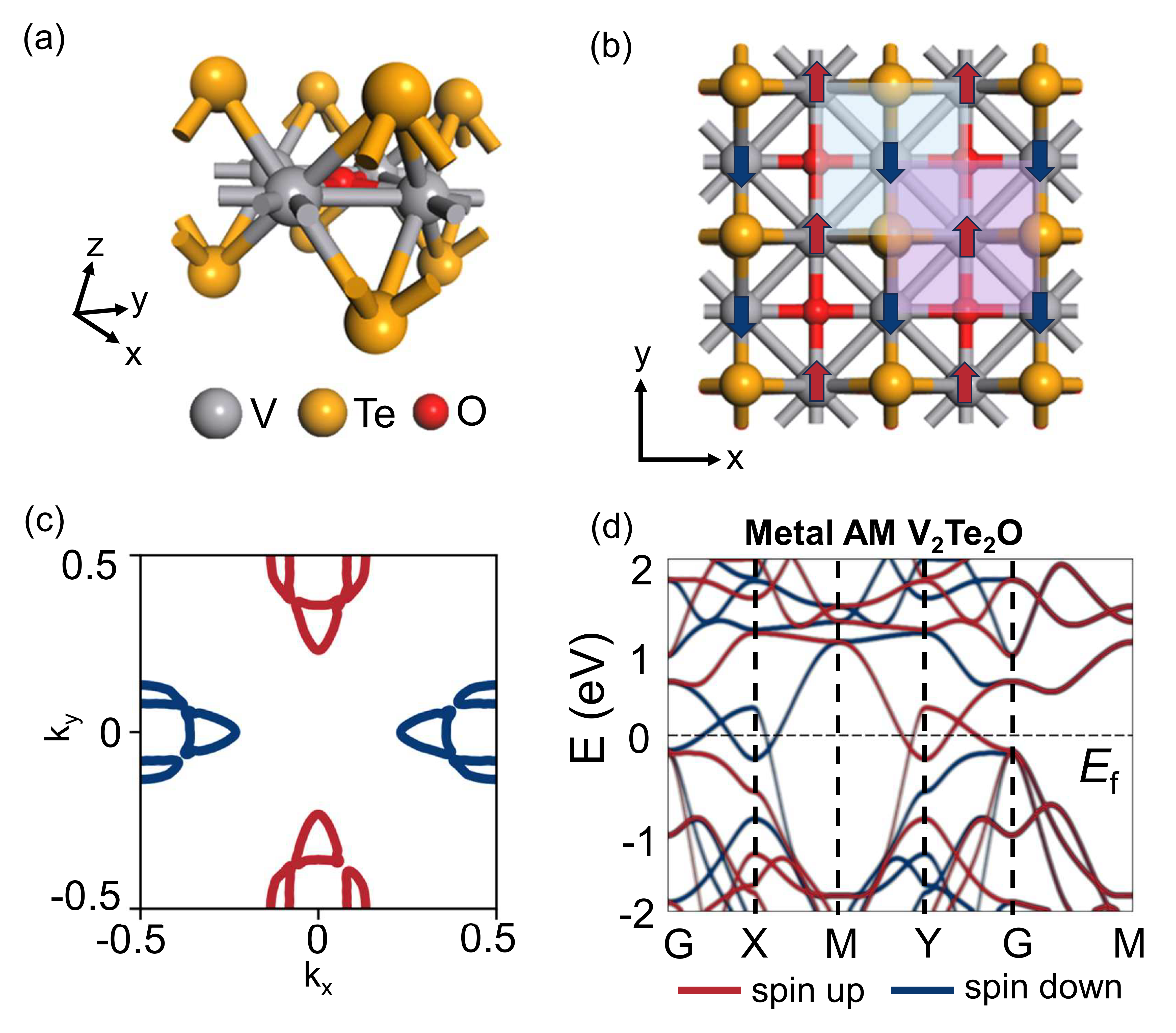}
  \caption{Structural and electronic properties of monolayer V$_2$Te$_2$O. (a) Crystal structure of V$_2$Te$_2$O, with V, Te, and O atoms shown in gray, yellow, and red, respectively. (b) Real-space magnetic configuration of the altermagnetic state. (c) Fermi surface in the $k_x$--$k_y$ plane, showing anisotropic spin splitting between spin-up (red) and spin-down (blue) channels. (d) Electronic band structure along high-symmetry paths, exhibiting momentum-dependent spin splitting characteristic of altermagnetism. The dashed line indicates the Fermi level.}
  \label{fig:fig2}
\end{figure}

As a representative material platform for the proposed device, we consider the two-dimensional Lieb-type altermagnet V$_2$Te$_2$O. The optimized lattice constant of V$_2$Te$_2$O is $a=b=4.02$~\AA. Its space group is $P4/nmm$, and it exhibits the characteristic $d$-wave altermagnetic spin splitting. As shown in Fig.~\ref{fig:fig2}, the material displays a real-space magnetic configuration compatible with altermagnetism, a spin-resolved Fermi surface, and momentum-dependent spin splitting in the band structure. In particular, electronic states along the $\Gamma$--X(Y)--M directions exhibit spin-up and spin-down polarization, respectively, consistent with the altermagnetic band texture.

V$_2$Te$_2$O is an appealing material platform for the present device concept for several reasons. First, it is predicted to host robust altermagnetic band splitting together with metallic transport characteristics, making it suitable as an electrode material in spin-photovoltaic junctions \cite{xu2026chemical}. Second, its bulk counterpart, Rb$_{1-\delta}$V$_2$Te$_2$O, has already been synthesized and confirmed as a room-temperature metallic compound \cite{Zhang2025SpinValleyLockingAltermagnet}, which supports the experimental plausibility of its two-dimensional counterpart. In addition, the bulk system shares the same Lieb-type lattice topology as the monolayer, suggesting that the essential altermagnetic electronic structure may be retained upon dimensional reduction. Finally, V$_2$Te$_2$O is theoretically predicted to possess a high N\'eel temperature of approximately 740 K \cite{Gong2024HiddenTopologyAntiferromagnets}, indicating that it may provide sufficiently robust magnetic order for practical spintronic and optoelectronic applications.

\begin{figure}[t]
  \centering
  \includegraphics[width=\columnwidth]{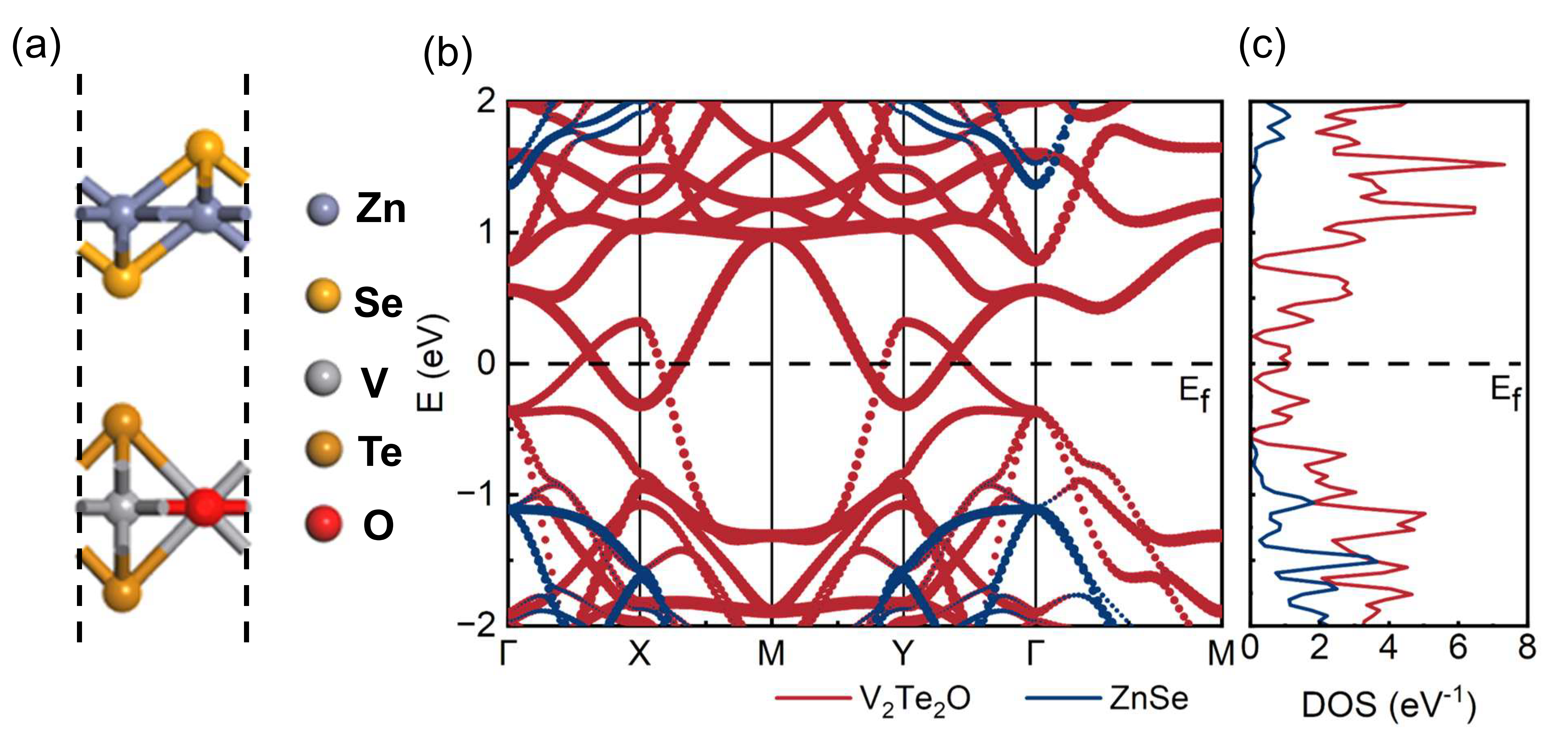}
  \caption{Electronic structure of the V$_2$Te$_2$O/ZnSe heterostructure. (a) Atomic structure of the interface, with Zn, Se, V, Te, and O atoms shown in blue, yellow, gray, orange, and red, respectively. (b) Projected band structure, where red and blue denote contributions from V$_2$Te$_2$O and ZnSe, respectively. (c) Corresponding projected density of states (DOS). The dashed line indicates the Fermi level.}
  \label{fig:fig3}
\end{figure}

To realize the proposed device concept, it is necessary to combine an altermagnetic electrode with a suitable semiconducting layer that can serve as the photoactive region while preserving spin-independent optical excitation. We find that ZnSe is a suitable candidate owing to its nonmagnetic character, moderate band gap, and comparable optical absorption \cite{xuan2022ferroelasticity,Yang2025TMRAltermagnets}. ZnSe is a nonmagnetic semiconductor crystallizing in the tetragonal $P4/nmm$ space group, with a band gap of 1.70~eV, as shown in the Fig.~S1 of the Supplemental Materials \cite{supp}. The optimized lattice constants of ZnSe are $a=b=4.11$~\AA. To construct the V$_2$Te$_2$O/ZnSe heterostructure, the ZnSe layer was slightly compressed to match the lattice of V$_2$Te$_2$O, resulting in a lattice mismatch of only 1.72\%, which indicates good structural compatibility at the interface. In addition, the optical absorption of monolayer ZnSe along the $x$ and $y$ directions nearly coincides, with a maximum value of about 19.25\% at $E=4.52$~eV, as shown in the Fig.~S2 of the Supplemental Materials \cite{supp}, suggesting that the semiconducting layer does not introduce spin-selective optical absorption by itself. Figure~\ref{fig:fig3} presents the atomic structure, projected band structure, and density of states of the V$_2$Te$_2$O/ZnSe heterostructure. The band alignment indicates the formation of a metal-semiconductor-metal junction, providing a suitable heterostructure platform for subsequent spin-selective photocarrier extraction.

\subsection{Quantum transport simulation}

\begin{figure}[t]
  \centering
  \includegraphics[width=\columnwidth]{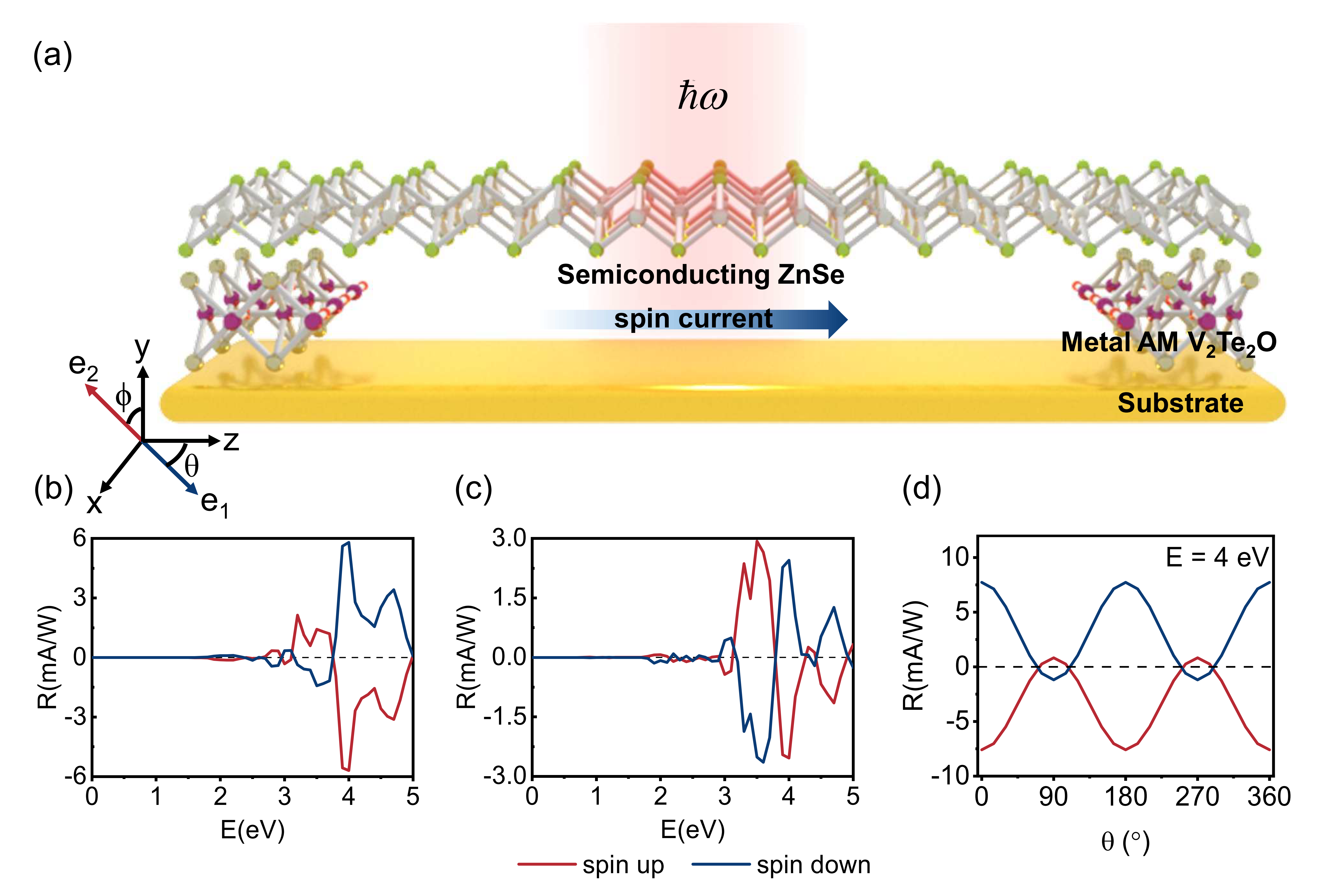}
  \caption{Photoresponse of the altermagnetic spin photovoltaic device. (a) Schematic illustration of the device under illumination, with a ZnSe semiconducting layer sandwiched between altermagnetic electrodes. (b,c) Spin-resolved photocurrent as a function of photon energy for parallel (P) and antiparallel (AP) N\'eel configurations. (d) Polarization angular dependence of the photocurrent at $E=4$ eV. Red and blue curves denote spin-up and spin-down photocurrents, respectively. $\theta$ represents the angle between the polarization direction and the vector $e_{1}$.}
  \label{fig:fig4}
\end{figure}

The designed altermagnetic spin photovoltaic battery adopts a lateral metal-semiconductor-metal geometry, in which a monolayer ZnSe strip is sandwiched between two monolayer metallic V$_2$Te$_2$O electrodes and placed on a substrate, as illustrated in Fig.~\ref{fig:fig4}(a). Unless otherwise specified, the device is considered in the AP configuration, in which the two altermagnetic electrodes possess opposite N\'eel vectors. Under illumination, the central ZnSe region serves as the photoactive semiconductor, while the two V$_2$Te$_2$O electrodes act as spin-selective contacts for carrier extraction. The photoexcitation occurs in the ZnSe layer, where spin-degenerate carriers are generated due to its nonmagnetic nature and are subsequently extracted through spin-selective channels imposed by the altermagnetic electrodes. The local density of device states (LDDOS) for spin-up and spin-down channels is shown in Fig.~S3 of the Supplemental Materials~\cite{supp}, where only a slight asymmetry between the two electrodes is observed, indicating negligible net spin polarization while retaining spin-dependent electronic structure, consistent with the altermagnetic character of the electrodes. The transport direction is aligned along the [100] axis, where the altermagnetic spin splitting is finite, whereas along the [110] direction the spin channels become degenerate, suppressing the spin-selective response, as confirmed by the nearly vanishing spin-resolved photocurrent shown in Fig.~S4 of the Supplemental Materials~\cite{supp}.

As shown in Fig.~\ref{fig:fig4}(a), the polarization of the incident light is described by the complex vector \(e=(\cos\phi\cos\theta-i\sin\phi\sin\theta)e_1+(\cos\phi\sin\theta-i\sin\phi\cos\theta)e_2\), where \(e_1\) and \(e_2\) are unit vectors specifying the polarization directions of the electric field, and \(\theta\) and \(\phi\) denote the polarization angles. The photoresponsivity $R$ is defined as the ratio of the photocurrent to the incident light power, $R=I_{ph}/P_{ph}$, where $I_{ph}$ and $P_{ph}$ represent the photocurrent and incident light power, respectively. The incident light power can be expressed as $P_{ph}=S\times n\times E$, where $S$, $n$, and $E$ correspond to the area of the central region of the device, the number of incident photons, and the energy of each photon, respectively. Under vertically incident linearly polarized light, corresponding to \(\theta = 90^\circ\) and \(\phi = 0\), the photoresponsivity is calculated as a function of photon energy, as shown in Fig.~\ref{fig:fig4}(b). In the antiparallel (AP) state, the spin-up and spin-down photocurrents flow toward opposite electrodes with nearly equal magnitudes and opposite signs, thereby generating a pure spin current. For example, at a photon energy of 4~eV, the spin-up and spin-down photoresponsivities reach $-5.69$ and $5.80$~mA/W, respectively. Notably, the onset of the photoresponse is consistent with the band gap of ZnSe ($\sim$1.70~eV), indicating that the photocurrent originates from optical excitation within the semiconducting layer. 

A similar behavior is observed under vertically incident circularly polarized light, corresponding to $\theta = 0$ and $\phi = \pi/4$, as shown in Fig.~\ref{fig:fig4}(c), where the spin-up and spin-down photocurrents remain nearly equal in magnitude and opposite in sign, with a maximum spin-up photoresponsivity of 2.93~mA/W and a corresponding spin-down value of $-2.51$~mA/W. Figure~\ref{fig:fig4}(d) further shows the angular dependence of the photoresponsivity under linearly polarized light at a fixed photon energy, corresponding to $\phi = 0$, where the spin-up and spin-down photocurrents exhibit an approximately cosine-like variation and remain nearly antisymmetric with respect to the horizontal axis. These results demonstrate that the pure spin photocurrent is robust against light polarization and exhibits a well-defined angular dependence, highlighting the intrinsic and symmetry-driven nature of the spin-selective photoresponse in the altermagnetic device. It is worth noting that, in the P state, the device does not exhibit equal-magnitude and opposite spin-polarized photocurrents, such that no pure spin current is generated, as shown in Fig.~S5 of the Supplemental Materials~\cite{supp}, which is consistent with Fig.~\ref{fig:fig1}(b).

\section{Conclusion}
In this work, we proposed an altermagnetic spin photovoltaic battery based on a V$_2$Te$_2$O/ZnSe/V$_2$Te$_2$O junction and investigated its spin-dependent photoresponse using first-principles quantum-transport simulations. We showed that when the two altermagnetic electrodes possessed opposite N\'eel vectors, the device supported a pure spin photocurrent. In this antiparallel configuration, the spin-up and spin-down photocurrents were equal in magnitude and opposite in sign, resulting in a vanishing charge photocurrent and a finite spin current. We further found that this effect persisted under both linearly and circularly polarized light and remained tunable with respect to photon energy and light polarization angle. These results establish altermagnetic junctions as a viable platform for light-driven pure spin-current generation and highlight their potential for altermagnetic spin-optoelectronic applications.

\begin{acknowledgments}
This work was supported by the High-Level Research Achievement Cultivation Project Funding of Hubei Minzu University (No. XN2321), the Educational Commission of Hubei Province of China (No. T201914), and the Science and Technology Innovation Project Funding for Youth Talent of Enshi (No. D20220066). Y.~S.~Ang was supported by the Singapore Ministry of Education (MOE) Academic Research Fund (AcRF) Tier 2 grant under Award No. MOE-T2EP50125-0019. Z.~Y.~Yang and X.~Y.~Yang were supported by the High-Performance Computing Platform of Peking University. L.~Z. was supported by the International Training Program for Doctoral Students of Northwest A\&F University (No. Z1050224003). H.~Z. was supported by the National Natural Science Foundation of China (Grant No. 12004307). Q.~L. and J.~W. acknowledged financial support from the China Scholarship Council (CSC).
\end{acknowledgments}

\bibliographystyle{apsrev4-2}
\bibliography{AM_References}

\end{document}